\RequirePackage{amsmath}
\documentclass[runningheads]{llncs}
\usepackage{hyperref}
\hypersetup{
    colorlinks=true,
    linkcolor=blue,
    filecolor=magenta,      
    urlcolor=cyan,
}
\usepackage{graphicx}
\usepackage{amsmath,amssymb} 
\usepackage{color}
\usepackage[caption=false]{subfig}
\DeclareMathOperator*{\argmax}{arg\,max}

\begin{document}
\title{Bayesian Inference-enabled Precise Optical Wavelength Estimation using Transition Metal Dichalcogenide Thin Films} 

\titlerunning{Bayesian Inference-enabled Precise Optical Wavelength Estimation}
%
\author{Davoud Hejazi$^1$, Shuangjun Liu$^2$, Sarah Ostadabbas$^2$, \and Swastik Kar$^1$}
%
\authorrunning{D. Hejazi, S. Liu, S. Ostadabbas, and S. Kar}
%

\institute{$^1$Department of Physics\\
$^2$Augmented Cognition Lab, Electrical and Computer Engineering Department\\
	Northeastern University, Boston, USA\\
}

\newcommand{\figgradelightT}{
\begin{figure*}[t]
\centering
\begin{tabular}{cc}
\hspace{-0.5in}
\includegraphics[width=0.75\linewidth, trim=0.0in 0.0in 0.0in 0.0in,
  clip=true]{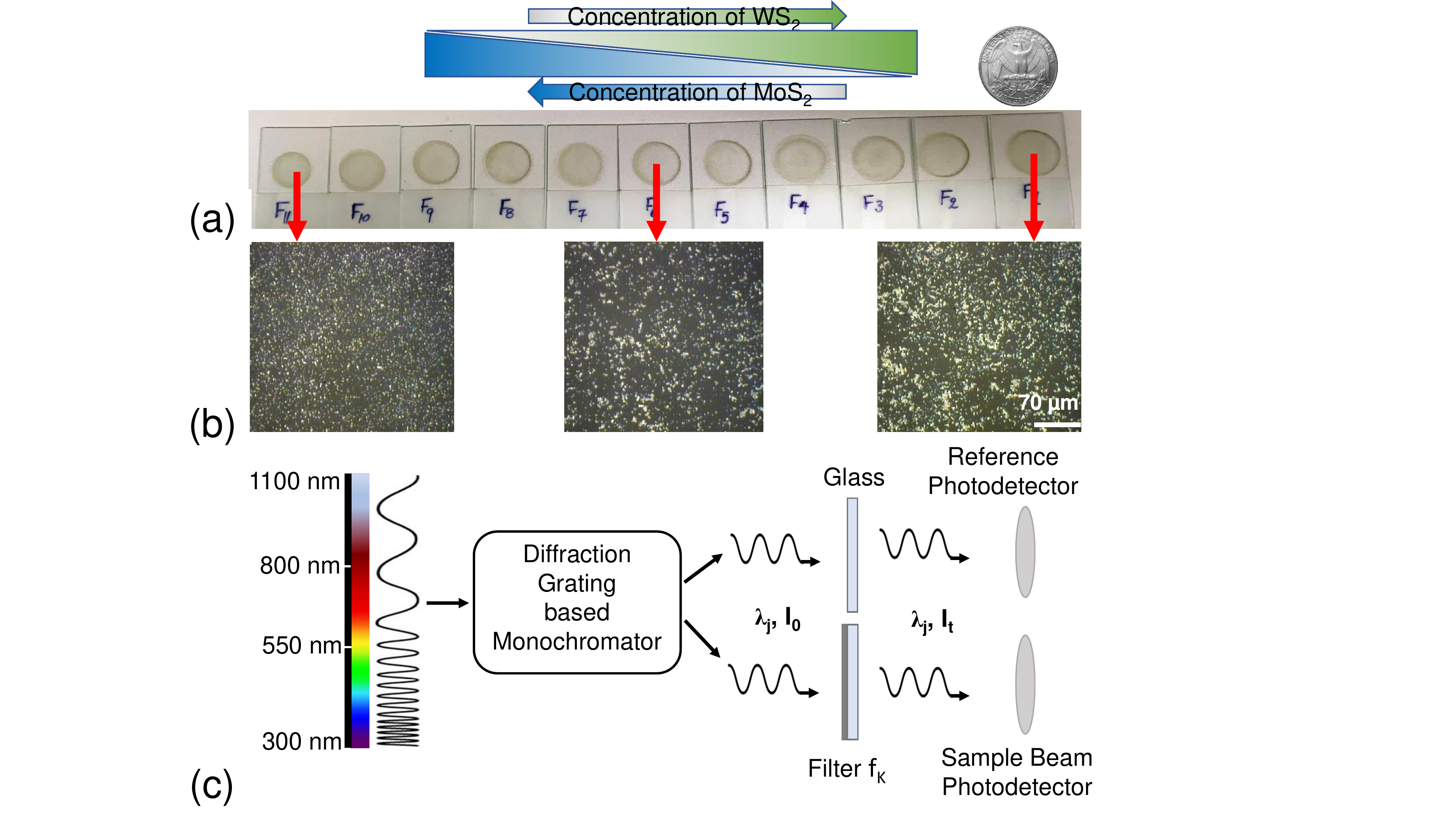}
&
\hspace{-0.9in}
\includegraphics[width=0.55\textwidth]{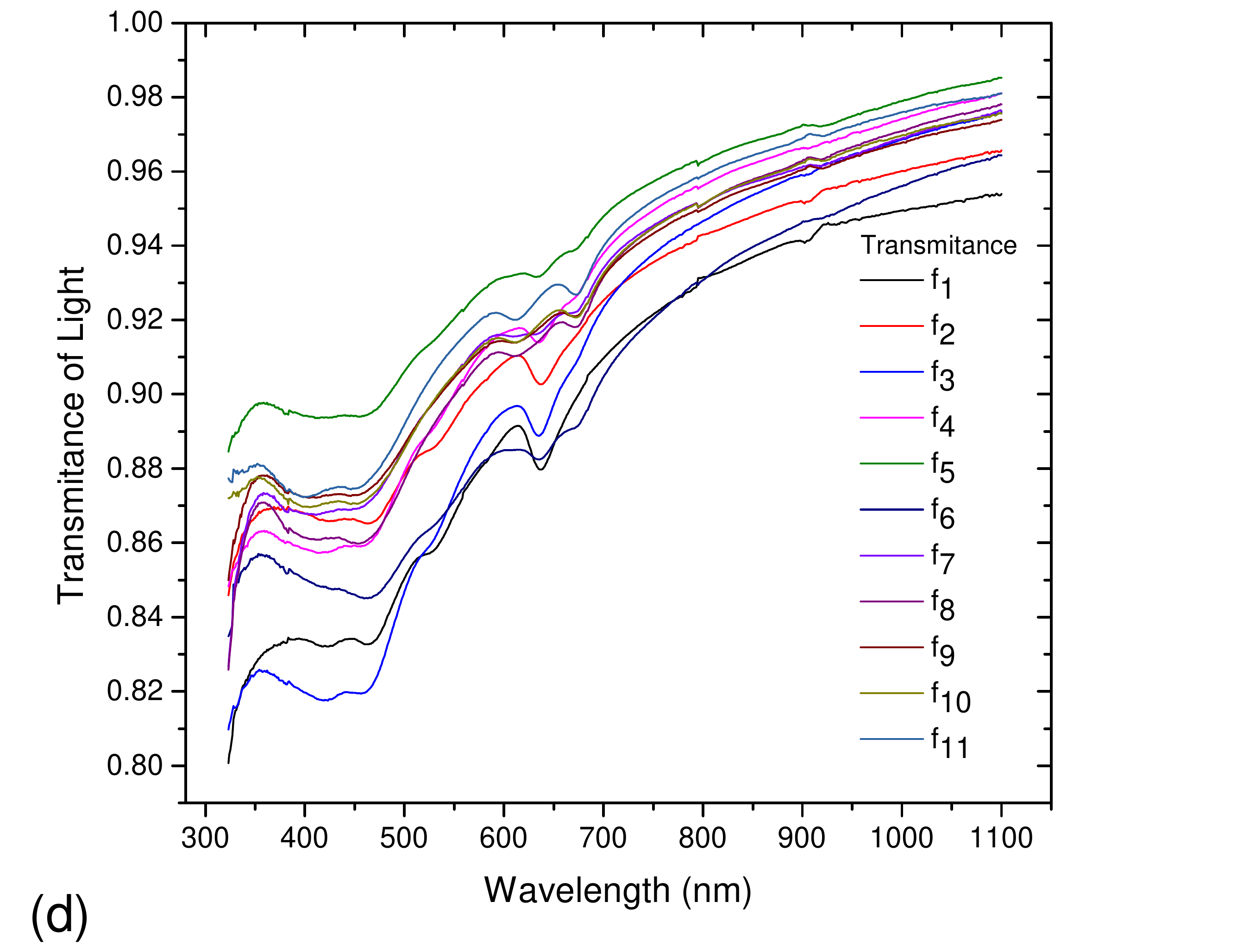}
\end{tabular}
\caption{(a) Eleven filters drop-casted on glass slide; $f_{1}$ is 100\% $WS_{2}$, but $f_{2},\dots,f_{10}$ are made by gradually adding $MoS_{2}$ and decreasing $WS_{2}$, and finally $f_{11}$ is 100\% $MoS_{2}$. (b) Microscopic image of three filters $f_{1}$, $f_{6}$, and $f_{11}$: nanomaterials on glass substrate. Atomic force microscopy (AFM) and scanning electron microscopy (SEM) images can be found in the Supporting Information. (c) Schematics of transmittance measurement as measured inside a Perkin-Elmer Lambda 35 UV-vis-NIR spectrometer. A broadband light source (a combiation of  deuterium and halogen lamps) that contains a spectrum of different wavelengths passes through a diffraction grating based monochromator. The monochromator isolates a narrow-band portion of the spectrum. This beam is split in two, passing through filter position and reference position, the beams are incident on photodetectors. A pure glass slide is placed in reference position and its spectral transmittance is removed from the total transmittance. (d) Background-subtraction transmittance $vs.$ wavelength for all 11 filters. The excitonic peaks get modified gradually from $f_{1}$ to $f_{11}$ as a results of changing proportion of mixing two TMDs.}
\label{fig:gradelightT}
\end{figure*}
}

\newcommand{\fighypothesis}{
\begin{figure}[t]
\centering
\includegraphics[width=1.04\linewidth, trim=0.0in 0.0in 0.0in 0.0in,
  clip=true]{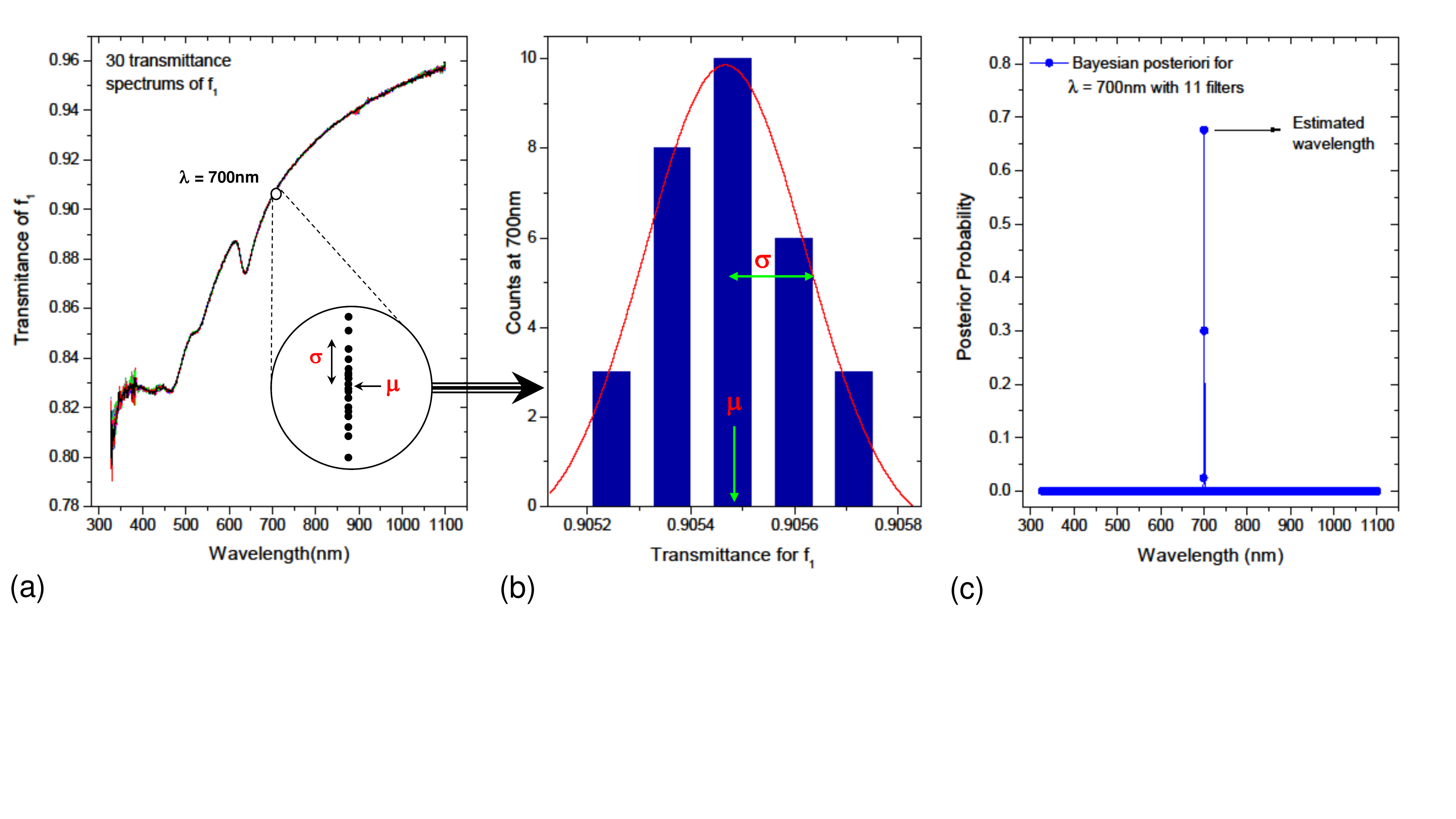}
\vspace{-0.9in}
\caption{(a) Transmittance spectra of $f_{1}$ measured 30 separated times, such each point on the curve is actually 30 dots as shown for 700nm in the inset. The mean value and standard deviation for each filter at each wavelength is calculated from this repeated training data. (b) Histogram of the same 30 transmittance data of $f_{1}$ at 700nm. The red curve shows a Gaussian function fit to the data. The parameters of the Gaussian distribution for each filter at each wavelength (\emph{i.e.} their $\mu$'s and $\sigma$'s) are found  using the training data. (c) The posterior probability calculated using Bayesian inference applied on transmittance data collected from our 11 filters when a test 700nm monochromatic light was shine on them.  The wavelength with maximum posterior is chosen as the estimated wavelength $\lambda^*$, which is equal to 700nm in this case.}
\label{fig:hypothesis}
\end{figure}
}

\newcommand{\figbayesianknn}{
\begin{figure*}[t]
\centering
\begin{tabular}{cc}
\includegraphics[width=0.57\linewidth, trim=0.0in 0.0in 0.0in 0.0in,
  clip=true]{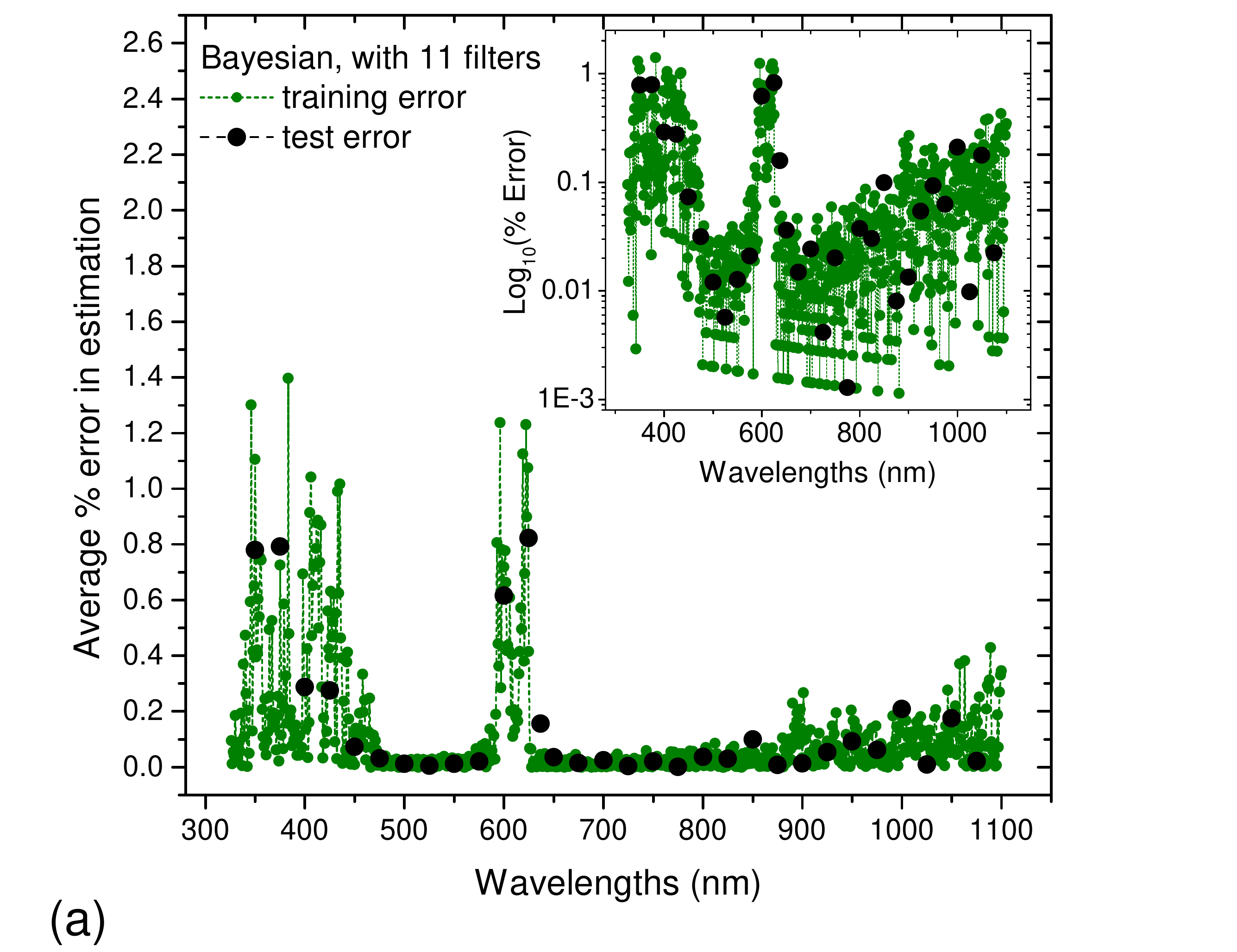}
 &
\hspace{-0.5in}
\includegraphics[width=0.57\textwidth]{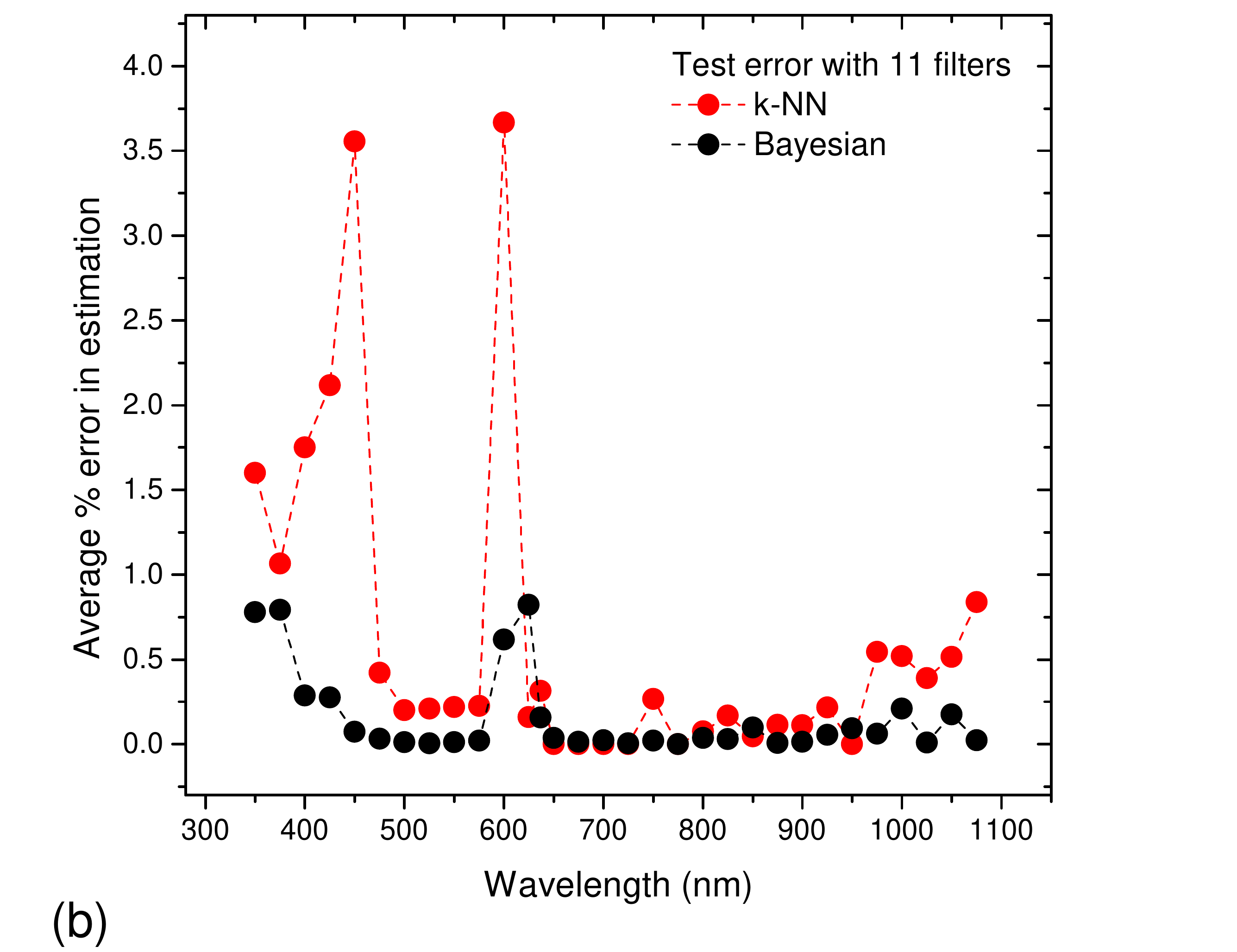} 
\end{tabular}
\caption{(a) Average training (in green) and test (in black) wavelength estimation error percent of Bayesian inference using all 11 filters. The inset is semi-log plot of the same figure. Each data point is averaged over 100 estimated values. (b) Average test wavelength estimation error percent by applying Bayesian inference (in black) compared to k-NN algorithm (in red) using 11 filters.}
\label{fig:bayesianknn}
\end{figure*}
}

\newcommand{\figentirebayesian}{
\begin{figure}[t]
\centering
\hspace{-0.1in}
\includegraphics[trim={0in 0in 0in 0},width=1.015\linewidth]{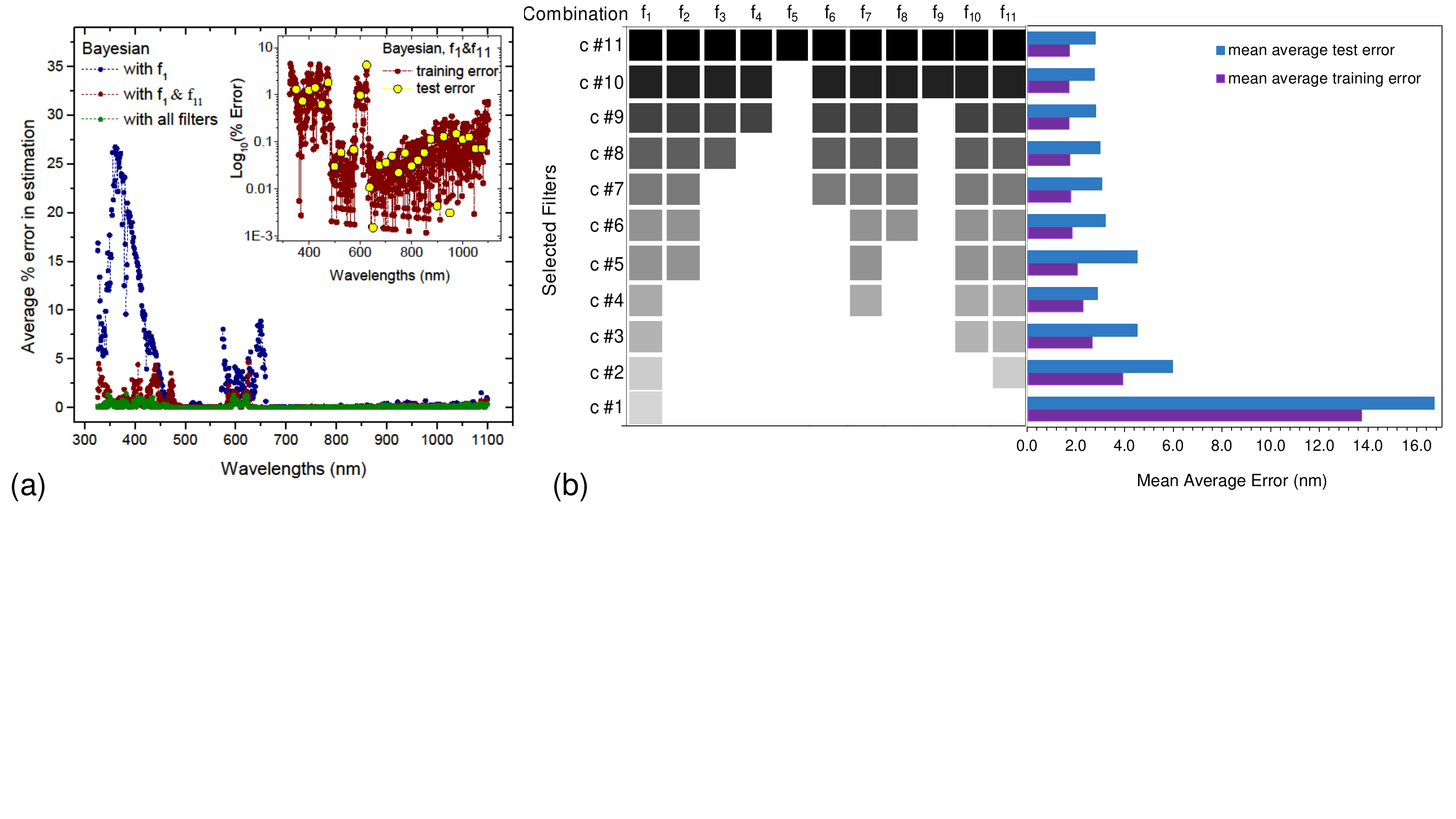}
\vspace{-1.2in}
\caption{(a) Average training error percent of Bayesian estimation using one filter ($f_{1}$), two filters ($f_{1}$ and $f_{11}$) and all filters (see text for definition of error percent). The inset semi-log plot of training and test percent errors when only two filters are used for estimation. Each data point is averaged over 100 estimations (see \secref{experiment}). (b) Filter selection via Greedy algorithm: horizontal dark-grey blocks represent the vector of chosen filters to use, and corresponding blue-purple pair of bars on the right show the mean average test or training error when the chosen filters are used for Bayesian estimation.}
\label{fig:entirebayesian}
\end{figure}
}

\newcommand{\figderivative}{
\begin{figure*}[t]
\centering
\begin{tabular}{cc}
\hspace{-0.15in}
\includegraphics[trim={0in 0in 0in 0},width=0.53\linewidth]{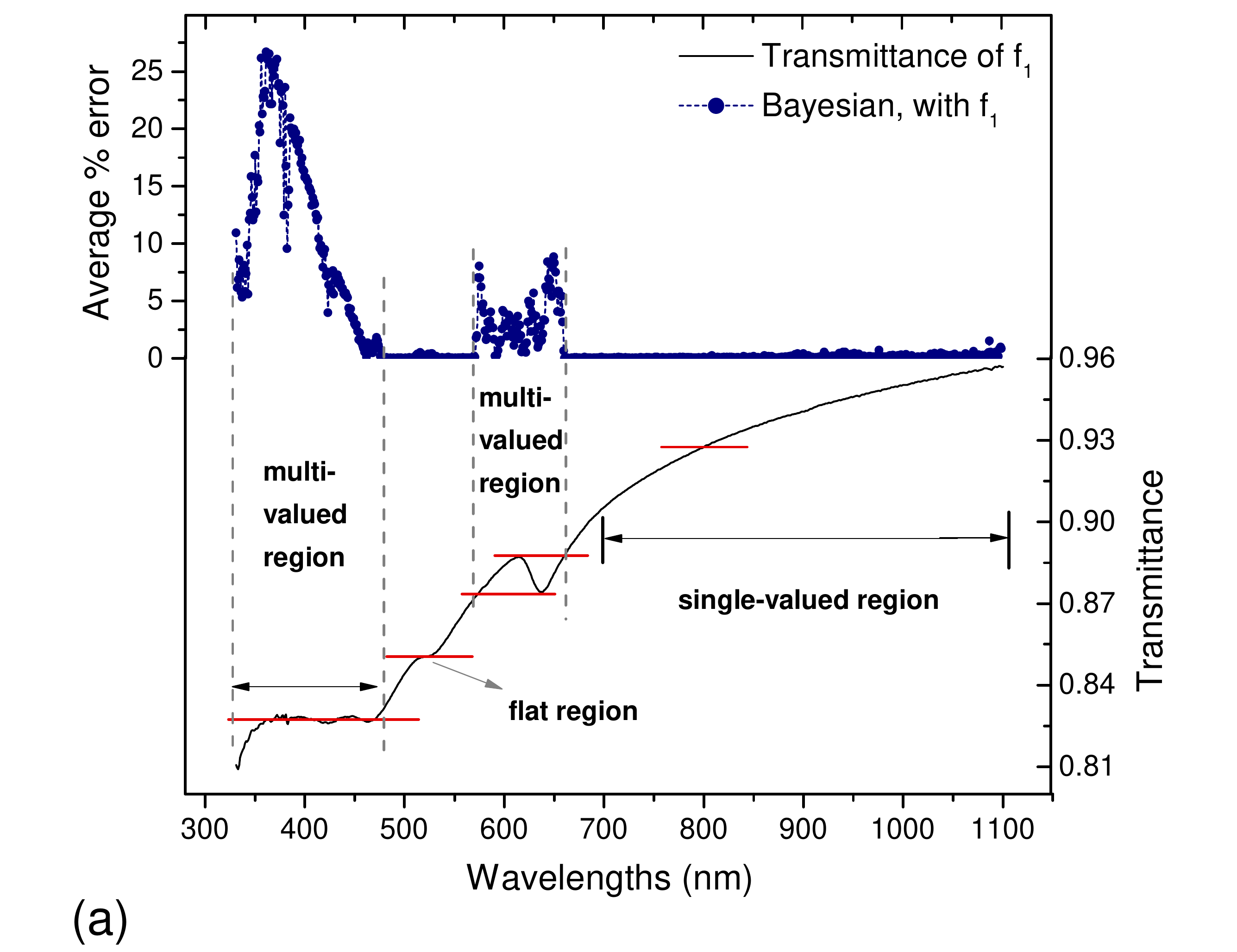}
\hspace{-0.25in}
\includegraphics[width=0.55\linewidth]{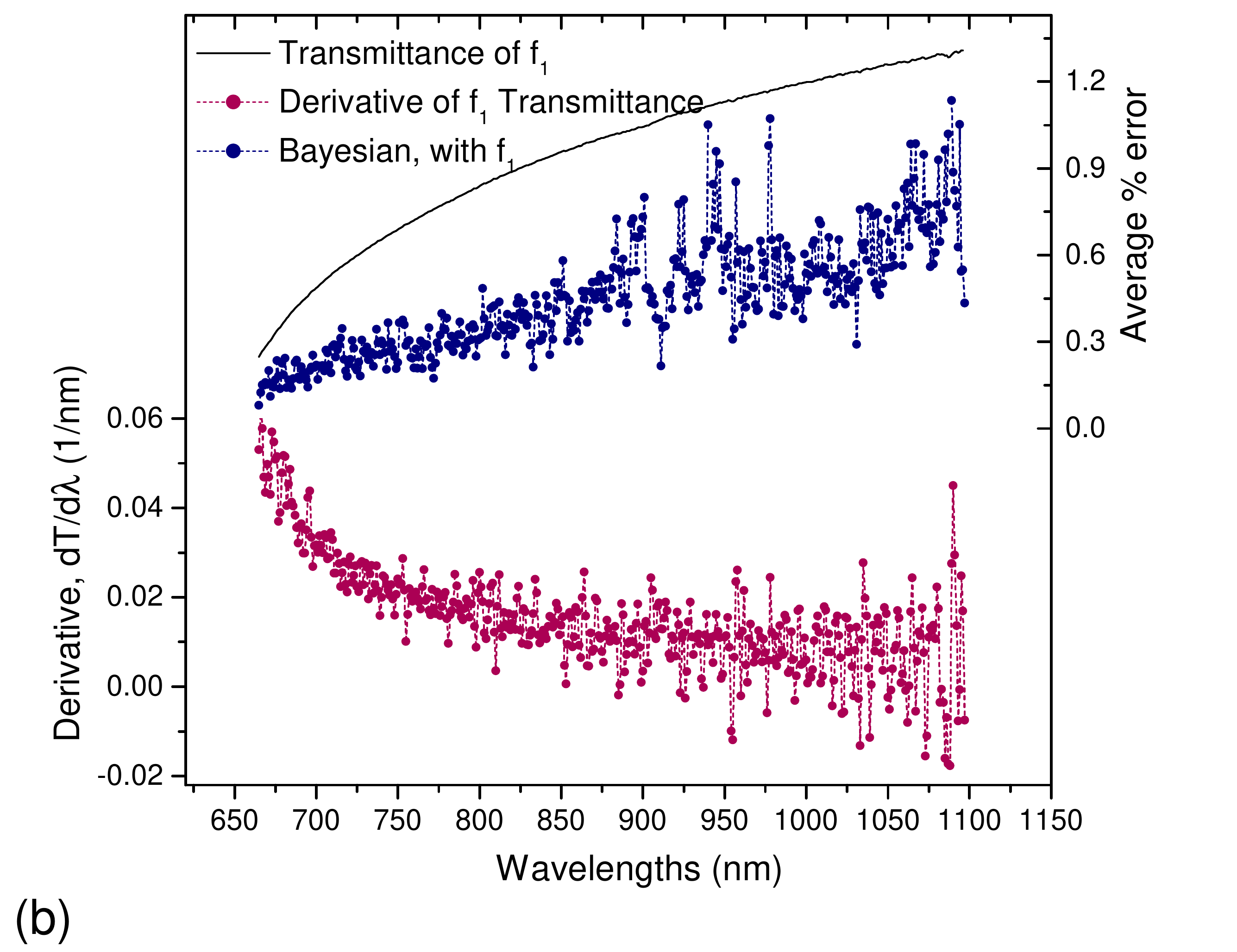}
\end{tabular}
\caption{(a) Average training error percent of Bayesian estimation when using only $f_{1}$ (in dark-blue) and transmittance of the same filter (black curve). Vertical dashed-lines indicate that co-existing different wavelengths with same transmittance values in the same neighborhood leads to inaccuracy in estimating wavelength; but when the transmittance is single-valued the error is small. The two multi-valued regions have obvious effects in increased error. Apart from the main single-valued region that is shown, two more of such regions exist between 500--600nm. The flat region causes only small increase in error. (b) Selected range of previous figure showing 665--1100nm that has monotonic (increasing only) transmittance (black curve). Also plotted in the same range is the first derivative (slope) of the transmittance $vs.$ wavelength (in blue) is the percentage estimation error (in pink). As the derivative of transmittance becomes smaller (decreasing slope), the errors become larger. For better visualization, the normalized RMS error\% is given in \figref{derivative}(b) (see Supporting Information for the equation).}
\label{fig:derivative}
\end{figure*}
}

\newcommand{\figoldnew}{
\begin{figure*}[t]
\centering
\begin{tabular}{cc}
\hspace{-0.15in}
\includegraphics[trim={0in 0in 0in 0},width=0.58\linewidth]{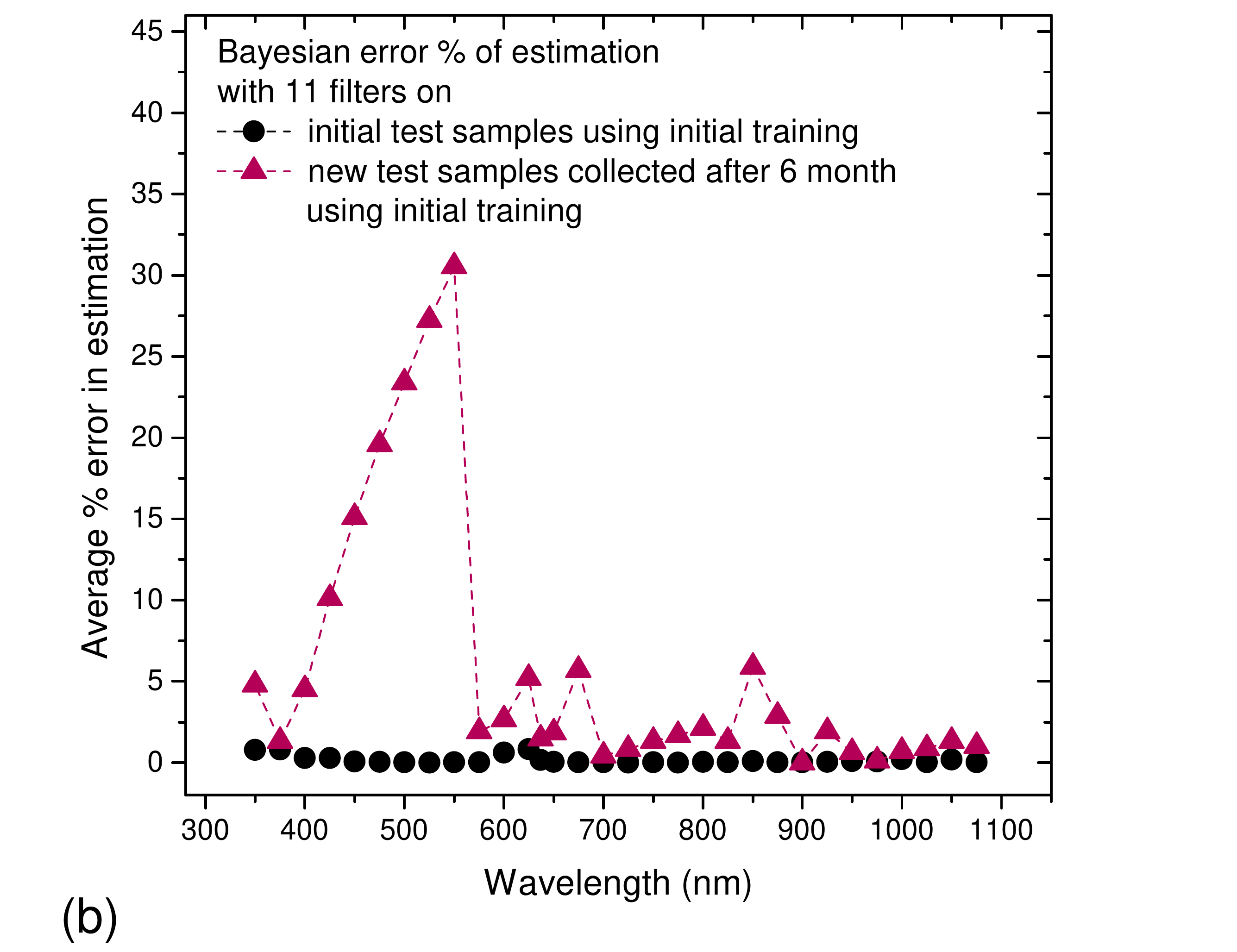}
\hspace{-0.5in}
\includegraphics[width=0.58\linewidth]{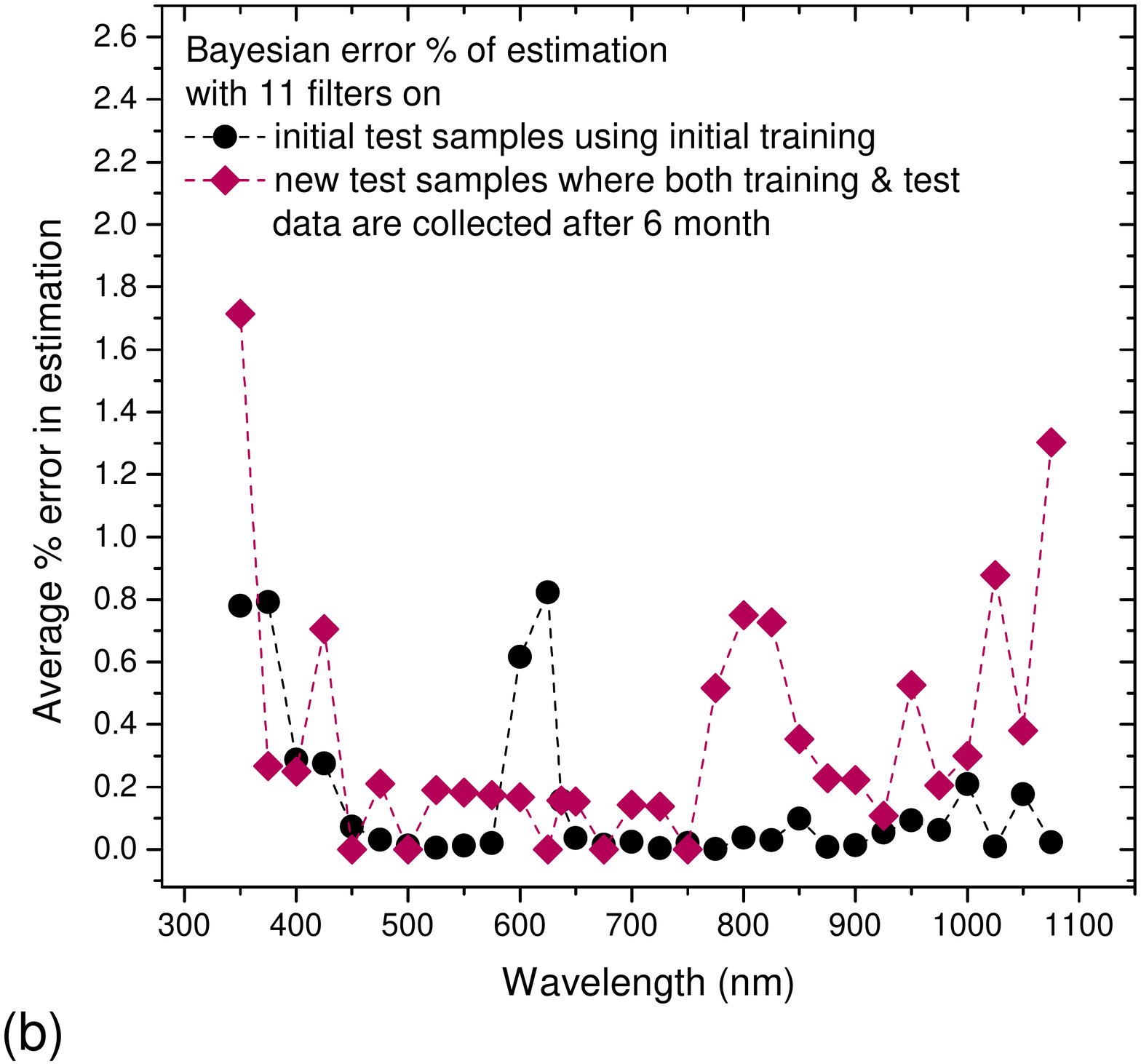}
\end{tabular}
\caption{(a) Average test error percent of Bayesian estimation with initial training on test samples collected at the same time (in black; same points as in \figref{bayesianknn}) compared to error on test samples collected after 6 month but still using the initial training set (in magenta). (b) Average test error percent of Bayesian estimation with the initial training data on test samples collected at the same time (in black) compared to the estimations with new calibration (6 month after the first calibration) using new training and test samples (all collected 6 month after first calibration - in magenta).}
\label{fig:oldnew}
\end{figure*}
}

\newcommand{\eqnref}[1]{Eq.~(\ref{eqn:#1})}
\newcommand{\figref}[1]{Fig.~\ref{fig:#1}}
\newcommand{\tblref}[1]{Table~\ref{tbl:#1}}
\newcommand{\secref}[1]{Section~\ref{sec:#1}}
\newcommand{\thmref}[1]{Theorem~\ref{thm:#1}}
\newcommand{\defref}[1]{Definition~\ref{definition:#1}}
\newcommand{\lemref}[1]{Lemma~\ref{lem:#1}}
\newcommand{\com}[1]{\textcolor{red}{#1}}
\maketitle

\begin{abstract}
Despite its ability to draw precise inferences from large and complex datasets, the use of data analytics in the field of condensed matter and materials sciences --where vast quantities of complex metrology data are regularly generated-- has remained surprisingly limited. Specifically, such approaches could dramatically reduce the engineering complexities of devices that directly exploit the physical properties of materials. Here, we present a cyber-physical system for accurately estimating the wavelength of any monochromatic light in the range of 325--1100nm, by applying Bayesian inference on the optical transmittance data from a few low-cost, easy-to-fabricate thin film "filters" of layered transition metal dichalcogenides (TMDs) such as MoS$_{2}$ and WS$_{2}$. Wavelengths of tested monochromatic light could be estimated with only 1\% estimation error over 99\% of the stated spectral range, with lowest error values reaching as low as a few ten parts per million (ppm) in a system with only eleven filters. By step-wise elimination of filters with the least contribution toward accuracy, mean estimation accuracy of $\sim$99\% could be obtained even in a two-filter system. Furthermore, we provide a statistical approach for selecting the best "filter" material for any intended spectral range based on the spectral variation of transmittance within the desired range of wavelengths. And finally, we demonstrate that calibrating the data-driven models for the filters from time to time overcomes the minor drifts in their transmittance values, which allows using the same filters indefinitely. This work not only enables the development of simple cyber-physical photodetectors with high accuracy color-estimation, but also provides a framework for developing similar cyber-physical systems with drastically reduced complexity.

\end{abstract}

\section{Introduction}
The ability to perform wavelength-selective photodetection has remained one of the most exciting areas of research in optoelectronics \cite{hu2018wavelength,de2018graphene}, owing to its applications in advanced photonic circuits and systems \cite{johnston2015optoelectronics}. In their simplest form, a photodetector is a light-sensitive semiconductor that operating in a conductive, diode, or transistor mode responds by generating a change in a measurable voltage or current to an incident light  \cite{xia2009ultrafast,tang2008nanometre}. Most conventional semiconductor photodetectors are broad-band, $i.e.$ they respond to a broad range of wavelengths, and hence are not intrinsically wavelength-selective. To achieve wavelength selectivity in photodetection, in addition to the traditional use of color pre-filters, various approaches have recently been proposed, such as the use of nanomaterials and nanostructures including quantum-dots \cite{konstantatos2006ultrasensitive}, photonics/plasmonics arrays \cite{kim2014graphene,grigorenko2012graphene}, and cavity-based-resonators \cite{lai1994design}, each with characteristic responses to specific wavelengths. These and other approaches have paved the way for a variety of tunable photodetectors capable of responding selectively to incident light with a specific wavelength \cite{goossen1994voltage,wood1985wavelength}. 

The situation is significantly more complex, however, when a detection system has to identify the wavelength of $any$ incident light (and not just a specific one). Such systems, that are capable of accurately discerning the wavelength of incident light, have immense relevance for applications such as bionic vision  \cite{wilke2010stimulation,wang2014bionic,lippke1981distribution}, robotic vision \cite{boyer1987color}, and various industrial light detection \cite{rieke2003detection,copeland2009color,mori2003ultraviolet,goudjil1998photochromic}, as well as astronomical and military applications \cite{kipping2011detection,nun2014supervised}. Typically, a wavelength estimating system ($e.g.$ in spectrometers) uses either a large number of photodetectors or an intricate diffraction-grating based monochromator coupled to one or two photodetectors to perform the task. To appreciate their complexity, let us consider a photodetection system that is required to estimate the wavelength of monochromatic light between the range of 325--1100nm. A number of notch/band-pass filters can be used to achieve this, depending on the desired resolution \cite{decker1978optical,shenoi2005introduction}. For example, if wavelength estimation is desired with 1nm accuracy, it will require a complex design with several hundred photodetectors with 1nm-width notch filters to achieve arbitrary wavelength identification. Beyond the design complexity of such a system, developing such narrow-width notch filters for each nanometer range could be a significant engineering challenge by itself. Alternately, the incident light could be diffracted off via a diffraction grating, but it will still require the same large number of detectors placed in an array to obtain the desired precision. Other approaches, such as rotating gratings, would lead to cumbrous addition of electronic and mechanical parts. In other words, achieving high accuracy wavelength selectivity using a purely mechanical set of detection systems can be complex, bulky, and expensive.  

We show that by using a few easy-to-fabricate nanomaterial-based broadband thin film filters, and harnessing sophisticated statistical approaches on large training datasets, it is possible to dramatically reduce the physical complexity of an accurate wavelength estimator. Nanomaterials, due to their diverse electronic and optical properties are constantly being explored and used for variety of low-cost, sensitive, and scalable photodetection technologies \cite{an2013tunable}. In this context, transition metal dichalcogenides (TMDs) are considered to be among the leading candidates in sensing applications. TMD monolayers are atomically thin semiconductors of the type $MX_{2}$, with $M$ a transition metal atom (Mo, W, etc.) and $X$ a chalcogen atom (S, Se, or Te). One layer of $M$ atoms is sandwiched between two layers of $X$ atoms \cite{wilson1969transition}. TMD monolayers of $MoS_{2}$, $WS_{2}$, $MoSe_{2}$, $WSe_{2}$, and $MoTe_{2}$ have a direct band gap, and can be used in electronics as transistors and in optics as emitters and detectors \cite{splendiani2010emerging,radisavljevic2011single,berg2017layer}. In this work, we used nanoscale TMDs  to develop thin film broadband optical "filters". Although these are not monolayer TMDs, their thin films provide wide regions of variation of optical transmittance, and the broadband optical responses from just a few filters turn out to be far more useful in wavelength identification than using a large array of "notch" filters, as justified next.


In addition to the previously discussed complexity and cost issues, all of the previously mentioned traditional methods share a common limitation:  Except for the readings corresponding to a specific filter or detector that senses the incident wavelength, the data from the rest of the detectors/measurements are usually discarded. This loss is an inefficient use of available data, especially since the appropriate use of data science provides ways to harness all available data to substantially increase the estimation accuracy from large datasets. Among the more powerful estimation algorithms is Bayesian inference, which is a theory in the field of statistics based on the Bayesian interpretation of the probability where probability expresses a degree of belief in an event, which can change as new information is gathered, rather than a fixed value based upon frequency or propensity \cite{bernardo2001bayesian,lee1989bayesian}. Bayesian inference uses Bayes' theorem to compute and update probabilities after obtaining new data. Bayes' theorem describes the conditional probability of an event based on the gathered data as well as prior information or beliefs about the event or conditions related to the event. Since Bayesian inference treats probability as a degree of belief, Bayes' theorem can directly assign a probability distribution to a parameter or set of parameters that quantifies the belief in them \cite{bernardo2001bayesian,lee1989bayesian}.

In this study, we have utilized the Bayesian inference approach to show that is possible to exploit the wavelength-dependence of broadband optical transmittance of simple thin film optical filters (built using low-cost, liquid-phase exfoliated TMDs) to accurately estimate the wavelength of any monochromatic color ($i.e.$ the midpoint of a very narrow-band light source with $\sim$1nm width, over a wide spectral range, 325nm$<\lambda<$1100nm). To the best of our knowledge, this is the first time the efficacy of such a powerful statistical analysis is being utilized to "train" a set of physical sensing systems to provide high accuracy estimation of "test" light sources, thereby developing the world's first cyber-physical monochromatic color estimator. By using Bayesian inference on optical transmittance data of up to 11 of such filters, the wavelengths of "test" monochromatic light could be estimated with less than 0.1\% estimation error for 71\% of the spectrum, and less than 1.5\% error for the rest of the spectrum. Furthermore, it is shown that even though using data from all available 11 filters yields to the smallest estimation errors in general, a Greedy selection algorithm \cite{cormen2001greedy} could be applied to reduce the number of filters and complexity of the system while keeping the estimation accuracy in an acceptable range. Our selection algorithm progressively discards the filters that have least contribution towards the accuracy of the wavelength estimation, hence achieves the lowest number of filters for a given acceptable accuracy. The proposed algorithm would allow in principle to perform filter selection on the desired specific spectrum ranges and find the optimal filter combinations that work well on those ranges, which might be different in other ranges.

A remarkable outcome of our investigation is how such high  accuracies could be achieved from a relatively simple photodetection system, as enabled by the advanced data analytics. The physical part of our design required only two types of TMDs ($MoSe_2$ and $WS_2$) to fabricate all of the 11 filters with sufficient filter-to-filter variations, while liquid-phase exfoliation technique (which was used to fabricate the thin film filters) can be considered to be one of the simplest fabrication approaches. Since the efficacy of the estimation was found to be driven by the monotonic, $i.e.$ single-valued nature of the spectral transmittance (or, "filter functions") and not on their specific values and other variations, their fabrication remained simple, scalable, and low-cost, requiring very little  process-control. Moreover, by re-calibrating these filters from time to time, it was possible to retain the high accuracy wavelength estimation ability over extended period of time. This aspect of our investigation addresses a huge challenge in many devices fabricated using nanomaterials, whose properties often degrade with time, rendering them impractical for real-world applications. In the following sections, we discuss how these systems were built and characterized, outline the statistical data analysis employed for wavelength estimation, and present details of the functional efficacy of our cyber-physical sensor.

\section{Results and Discussion}
\label{sec:results}
\textbf{Filter Design and Transmittance Data.} The estimation of wavelengths was done by fabricating a set of eleven thin film optical filters using a combination of two TMDs nanoflakes, in order to get sufficient variations in their filter functions. \figref{gradelightT}(a,b) show digital images of all the filters, including microscope images from three representative filters. Two out of the eleven filers (which we labeled as $f_{1}$ and $f_{11}$) were made by drop-casting suspensions of liquid-phase exfoliated nanoflakes of the two types of TMDs, $MoS_{2}$ and $WS_{2}$, respectively, onto the surface of two separate glass slides. The other nine filters were made by drop-casting suspensions with gradually differing mixing proportions of the same two TMDs onto surface of separate glass slides. The transmittance $vs.$ wavelength of these filters over the 325--1100nm spectrum range was collected 120 times for each filter, using a Perkin-Elmer Lambda 35 UV-vis-NIR spectrophotometer. In this process, a broadband light source was converted into variable monochromatic light using a diffraction grating system and was made to pass through the filters, and the transmitted light was measured using a silicon photodetector. \figref{gradelightT}(c) schematically represents the apparatus measuring the  transmittance, where a plain glass slide was used to remove the background transmittance of the slides. The mean value of the glass-background-subtracted transmittance for all of the filters are shown in \figref{gradelightT}(d). In each case, the overall transmittance values were found to grow with increasing wavelength of incident light corresponding to the gradually reducing density of states near the Fermi level of these materials with growing wavelength (or decreasing energy values) tending towards zero close to the band gap. In addition, there are characteristic "dip" features that correspond to various excitonic resonances in these systems \cite{knox1963theory}. Mixing two different TMDs in different amounts enabled us to get gradually evolving transmittance curves with changing magnitudes, slopes, and feature positions. As we show, this allowed us to examine which features of the transmittance curve were responsible for higher wavelength estimation confidences. We next discuss how the data obtained using these filters were analyzed using the Bayesian inference.

\figgradelightT

\vspace{0.1in}
\textbf{Wavelength Estimation Using Bayesian Inference}: The statistical analysis of our data were performed over a set of transmittance values measured discretely over the entire range of wavelengths, for each filter, as well as 120 repetitions of wavelength-dependent data. The repeated data was acquired to account for drifts, fluctuations, and other variations commonly observed in physical measurements especially in nanomaterial-based systems, which tend to be sensitive to their environments. Using this data referred to as our "training data", we formulated the wavelength estimation problem as follows: Let $\lambda = \{\lambda_{1},..., \lambda_{i},...,\lambda_{N}\}$ be $N$ different wavelengths in desired spectral range and with specified granularity (\emph{i.e.} 325--1100nm with 1nm step in this study), and $T=\{t_{1},...,t_{i},...,t_{K}\}$ be the transmittance vector of $K$ filter values (\emph{i.e.} $K=11$ when all of the filters are used in this study). Employing the Bayesian inference, the probability of the monochromatic light having the wavelength $\lambda_{j}$ based on the observed/recorded transmittance vector $T$ data is called \emph{posterior} probability $P(\lambda_{j} \mid T)$, which is the probability of a hypothesis given the observed evidence: 

\begin{equation}
    P(\lambda_{j} \mid T)=\frac{P(T\mid \lambda_{j})P(\lambda_{j})}{P(T)},
\end{equation}
where $P(\lambda_{j})$ is the \emph{prior} probability of the specific wavelength $\lambda_{j}$ being present in the monochromatic light of interest, which is defined as the estimate of the probability of the hypothesis before the current evidence is observed. In this study, we assumed all of the wavelengths are equally-likely to happen, so we considered a uniform distribution function for the prior probability as $P(\lambda_{j})=\frac{1}{N}$, where $N$ is the total number of quantifiable wavelengths in the range under study. Moreover, $P(T\mid \lambda_{j})$ is the probability of observing transmittance data $T$ given wavelength $\lambda_{j}$, and is called the \emph{likelihood}, which indicates the compatibility of the evidence with the given hypothesis. Although, the filters are related due to having the same two materials with different mixtures, for computational purpose, we assume independence between their outcomes, and model them using
with Naive Bayes algorithm \cite{sahami1996learning}. As such, the likelihood of all filter readings $T$ can be calculated as the product of each filter value $t_i$ in a given wavelength $\lambda_{j}$, as $P(T\mid \lambda_{j})=\prod_{k=1}^{K}{P(t_{k}\mid \lambda_{j})}$. To compute individual $P(t_{k}\mid \lambda_{j})$ values, a Gaussian normal distribution for each filter at each wavelength was assumed, and their mean values and standard deviations were calculated from the training data ($i.e.$ the 120 measured transmittance spectra) collected from each filter at each wavelength. $P(T)$ is called \emph{marginal} probability of measured transmittance vector $T$, which can be calculated as $P(T)=\sum_{i=1}^{N}{P(T\mid \lambda_{i})P(\lambda_{i})}$. Since $P(T)$ is the same for all possible hypotheses that are being considered, so acts as a normalization factor to keep the posterior probability in the range of 0 to 1. Finally, given the measured transmittance sample $T$ (a vector of $K$ elements -- one transmittance value per filter at an unknown wavelength, see \secref{experiment} for more details), the target wavelength $\lambda^*$ of the monochromatic light is estimated by choosing the value of $\lambda_j$ that maximizes the posterior probability $P(\lambda_{j} \mid T)$:

\begin{equation}
   \lambda^* =  \argmax_{\lambda_{j}} P(\lambda_{j} \mid T),
\end{equation}
in which this optimisation called the maximum $a$ $posteriori$ (MAP) estimation \cite{bassett2016maximum,bernardo2001bayesian,lee1989bayesian}.

The efficacy of our wavelength estimator was tested both using test samples, $i.e$  transmittance value for a test monochromatic source that were collected separately, and which were not used in the training data and hence were not seen by the model before; as well as for training samples which were generated randomly from the same Gaussian distributions that were assigned to each wavelength for each filter. The training samples were utilized to check how well the model works on the training set itself, while the test samples are used to investigate how the model can estimate the truly unknown wavelengths. \figref{hypothesis} provides an typical example of applying Bayesian inference for estimating the wavelengths. The 
\figref{hypothesis}(a) plots 30 instances of overlapping transmittance spectra of the filter $f_1$, with the inset showing a magnified view of the transmittance data set for a single wavelength (shown here for $\lambda$ = 700 nm), the collected transmittance values creating a distribution around the mean value of transmittance for that wavelength. \figref{hypothesis}(b) shows the histogram of the same 30 recorded transmittance data of $f_1$ at 700nm with the calculated mean $\mu$ and standard deviation $\sigma$. The red curve shows the Gaussian fit on the data, which justifies the assumption of a normal distribution for the $P(t_{k}\mid \lambda_{j})$ probabilities. In order to perform Bayesian inference for wavelength estimation of a test monochromatic light, we needed to calculate the posterior probabilities of different wavelengths $P(\lambda_{j} \mid T)$, when transmittance data $T$ is collected from our 11 filters. When testing the efficacy of our wavelength estimator, a new "test transmittance data" set is collected separately from the training transmittance data. \figref{hypothesis}(c) shows the posterior probability as a function of wavelength. It is apparent that the maximum posterior probability is close to 1  around 700nm, and is almost zero for the rest of the spectrum, which indicates the reliability of the Bayesian inference. Several other cases are presented in Supporting Information. The same procedure was performed to estimate all of the wavelengths (test and training).

\fighypothesis

\figbayesianknn

\textbf{Wavelength Estimation Accuracy} To discuss the efficacy of our wavelength estimator, we define the estimation error as difference between average estimated wavelength and real wavelength, divided by the real wavelength, times 100 to find average estimation error percentage. The percentage average estimation error (when using all 11 filters) is plotted as a function of wavelength in \figref{bayesianknn}(a). In this figure, we plot the estimation error from both the "training" (in green) as well as the "test" (in black) data. We see that using only eleven filters and two photodiodes, our cyber-physical color estimator is able to achieve extremely high accuracy color-estimation, with just a very few estimation data points lying above 1$\%$ error, with most of the error values being far lower in comparison. To appreciate how high the accuracy values reached, the semi-log plot of the same data has been shown in the inset. We find that not only a significant portion of the estimated wavelengths from the training dataset was better than 0.1$\%$, the lowest errors are arrive close to 0.001\%, or a few tens parts per million. In particular, the test data, which could be performed only on a smaller set of source wavelengths, appear to fall well-within these low-error accuracies. This fact that errors for the estimation of the test samples are well-within the range of training errors, highlights an extremely important feature of our cyber-physical wavelength estimator, $i.e.$ the efficacy of our Bayesian inference approach is generalizable from training set to test data. In other words, a lab-trained system is very likely to continue providing high-confidence wavelength estimations under field-testing as well. We conclude that our cyber-physical approach using 11 filters and the Bayesian inference system is not only able to estimate unknown wavelengths with a high degree of accuracy, but also do so with equal efficacy under both training and testing conditions.

In order to compare the estimation results of our Bayesian inference with another data-driven approach, we chose 
the k-nearest neighbor (k-NN) as one of the most straightforward machine learning algorithm, which is widely used in supervised classification applications \cite{altman1992introduction}. For training and testing the k-NN, we used the same training and testing datasets we used in our Bayesian inference, respectively. Using the k-NN method, the wavelengths that had nearest transmittance values to the test samples were found, and the result for test data from all 11 filters has been plotted in \figref{bayesianknn}(b). It is noted that while in some instances, the k-NN approach provides estimations of nearly the same accuracy as the Bayesian approach, overall the estimation accuracy with the Bayesian methods is superior over the entire spectral range, especially at the lower wavelength values where the deviation of estimation from the real values are larger. For this reason, we present Bayesian inference as the the primary analysis approach for the rest of this study. We next investigate, in a step-wise manner, how the wavelength estimation efficacy changes as the number of filters are reduced. 

\figentirebayesian

\vspace{0.1in}
\textbf{Filter Selection and its Effect on Estimation Accuracy.} A key advantage of our cyber-physical system is that its ultimate estimation accuracy depends on both the efficacy of the Bayesian inference approach as well as the total number of filters used. In other words, if such high accuracy is not required for any specific application, it is possible to further reduce the physical complexity of the system. With all the training data available to us, it was possible to investigate the estimation accuracy of the system by identifying and removing the filters that were least effective, in a step-by-step manner. Understandably, by using a fewer number of filters for estimation, the error tends to increase. The estimation error $vs.$ wavelength when using only 1, 2, or all 11 filters are shown in \figref{entirebayesian}(a). We found out that using only two or one filter(s), the highest estimation errors grows by a factor of $\sim$5 and $\sim$25, respectively, at the most error-prone region between 300--500nm. In most of the remaining parts of the spectrum, the estimation error remains much lower, as seen more clearly in the the inset that shows a semi-log plot of the test and training error when using filters $f_{1}$ and $f_{11}$. Further, starting with all 11 filter functions, using the Greedy algorithm \cite{cormen2001greedy} as a filter selection approach, the number of filters could be sequentially reduced, by discarding the filters with least contribution towards accuracy of estimation one by one. This way, the complexity of the system could be reduced systematically, while minimizing the cost of reducing overall accuracy at each step. \figref{entirebayesian}(b) represents the results of this filter selection. Here, on the left side of the representation, each cell represents a filter function being used (gray box) or discarded (white space). Starting from the top, where all the filters are present, the Greedy algorithm was used to drop the least useful filter and this is represented in the next row. In this way, in each row, the least useful filter is dropped and the mean value of the average error (from the entire spectrum) is plotted as a horizontal bar-graph on the right end of the row, for both training and test errors. We found out the very encouraging result that the mean average error (presented in nanometers instead of percentage values) does not change much until it gets to the last few filters, suggesting that the filter-to-filter variation of transmittance functions using our simple approach provides was quite effective. Indeed, even with two filters the error is significantly small, with an effective average error of $\sim$6nm, which reflects less than $\sim$1$\%$ error at the center of the spectrum. Applying this feature selection method when only two filters are desired reveals that the filters $f_{1}$ and $f_{11}$ would give the best wavelength estimation results, which was expected because these two were the most independent filters being fabricated using completely independent nanomaterials, while the other filters are mixtures of both materials. The increase in error is more obvious when switching from 2 to 1 filter, which establishes that a single-filter photodetector would not be enough for reliable wavelength estimation.


\figderivative

\vspace{0.1in}
\textbf{Sources of Estimation Error.} We next discuss factors that affect the accuracy of estimation as related to the curve-shapes of the transmission functions. There is an interesting correlation between the positions (wavelength values) of local maxima/minima of transmittance curves (which arise from variations of the density of state and presence of excitonic peaks, fairly well-known features of the spectral absorption curves of TMDs \cite{knox1963theory}), as seen in \figref{gradelightT}(d)), and where the errors tend to increase. Large error occurs across multi-valued regions of transmittance curve, $i.e.$ regions where multiple wavelengths may have same or very similar transmittance values. The estimations using only one filter is shown in \figref{derivative}(a) for simplicity. The horizontal lines between two vertical dashed lines clearly show when the horizontal lines cut more than once through the transmittance curve, the error becomes larger; but when a horizontal line passes through only one point, the error is smaller. This result was expected because multiple wavelength will presents similar posterior values in predictions, which are hence prone to wrong estimations. Thus, materials with monotonic responses ($e.g.$ without excitonic or other absorption peaks, or in other words with single-valued spectral transmittances) are better choices for fabricating such filters. Furthermore, even in the parts of the spectrum that filter function is monotonic (in this case only increasing) the errors are smaller when the slope (derivative) of the transmittance curve is larger. \figref{derivative}(b) which refers only to a part of the spectrum between 665--1100nm reveals that as the slope (derivative) of transmittance decreases the error increases. For better capturing the the deviations in estimation visualizing the errors, the root-mean-square  (RMS) \% error is used here (see Supporting Information for the equation). From these results, we conclude that ideal transmittance curves should be monotonic with adequately changing transmittance values. We note that while conceptually this is not difficult to understand, in a real-world situation, it is challenging to "pre-order" the transmittance curves of any material, once again pointing towards the usefulness of the characteristic transmittance of the TMDs used in this study.  

\figoldnew

\vspace{0.1in}
\textbf{Filter Stability and Reusability Over Time.} Finally, we explored the estimation reproducibility of these easy-to-fabricate physical filters, which would be an extremely important consideration from a practical viewpoint. These filters were simply drop-casted onto the surface of regular glass without any additional protection, and the typical time-lapse between first calibration of filters estimation was 1$\sim$100 days, which demonstrates the physical stability of the filters despite being left in ambient conditions for 10\% of the time and under nominal vacuum for 90\% of the time. Still, gradual change of the optical properties in these nanomaterials is expected, as they absorb various gaseous species from the ambient. To check the stability of the filters, six months after the first calibration, a new test set was collected and was estimated using the original 6 month old training data. It was interesting to see that while the estimation errors were found to have become larger for smaller wavelengths values, but for most of the higher-wavelength portion of the spectrum the estimation error remained better than 3$\%$ as seen in \figref{oldnew}(a). 

To see whether the filters are reusable in longer time spans or not, a new training set also was collected along with the new test set (6 month after original calibration). Despite some minor changes observable in transmittance (optical response or filter function) of the filters, by calibrating the filters using the new set of training data it was possible to estimate the new unknown wavelength as accurate as before (\figref{oldnew}(b)). This, not only suggests fair stability of these nanomaterial filters but also shows that by calibrating the filters from time to time, it would be possible to continue using these same filters over extended periods of time, and the efficacy of estimations does not suffer from wear or minor scratches, since the calibration will overcome the gradual changes of the filters.

\section{Conclusions}
\label{sec:conclusion}

In conclusion, we have successfully demonstrated a new approach that applies data analytics ($i.e.$ Bayes's theorem) to the optical transmittance information collected from two extremely low-cost nanomaterial filters to estimate the wavelength of a narrow-band unknown incident light with high accuracy. Using more number of filters that are created from the same two nanomaterials it is possible to considerably improve the accuracy of estimation, and with a feature selection algorithm the minimum number of filters needed for an acceptable value of average accuracy can be determined by retaining the only the "most relevant" filters. Even though the experiment was performed over the range 325--1100nm, in principle this approach can be extended beyond in both the UV as well as NIR directions, thereby opening up the possibility of developing next generation wavelength-estimators for both visible as well as beyond-visible regions of the EM spectrum. The filters performed robustly even after many months without additional protection and only low-maintenance storage, and by re-calibrating the Bayesian inference model used for estimation from time to time it is possible to continue using these same filters with high accuracy over extended periods of time. In the ranges of spectrum that filter function (transmittance) has a monotonic dependence on wavelength, the estimation accuracy is higher, and furthermore, there is a positive correspondence between slope of filter function and accuracy of estimation. Hence, based on the application and desired spectral range, the highest accuracy values will be obtained by using materials (either TMDs or other transparent films) whose transmittance values show large but monotonic changes with wavelength. In addition to the Bayesian approach, the k-NN analysis was also successfully applied, though with comparatively lower wavelength estimation efficacy. We believe that our findings open up a completely new path for designing next generation sensors and detectors that can harness the power of data analytics to reduce the physical complexity of detectors in general, and in particular for future works on generic non-monochromatic lights using more advanced data analyzing methods and state-of-the-art machine learning techniques. We believe that this significantly transforms the field of high-accuracy sensing and detection using simple cyber-physical approaches.

\section{Materials and Methods}
\label{sec:experiment}

\textbf{Sonication-Assisted Liquid-Phase Exfoliation.} Bulk $MoS_{2}$ and powder of $WS_{2}$ were purchased from ACS material. Bulk $MoS_{2}$ was grinded using pestle and mortar, but the powder of $WS_{2}$ was used as received. 80mg of $MoS_{2}$ powder and 8mL of Isopropyl alcohol (IPA) was added into a beaker and stirred until the dispersion became dark, then was exfoliated in liquid phase via sonicating using 30kHz and 80\% of the power of a UP100H Hielscher ultrasound processor for 8hours while the beaker was placed in cool water to avoid overheating all the time. Afterwards, the dispersion was left still for a minute for the bulk materials to settle down; supernatant (top half) of the dispersion was collected and centrifuged for 2 minutes with 1000rpm using a Thermo Scientific centrifuge. The supernatant (top one-third) of the centrifuged dispersion was moved to another container and was centrifuged again with 1000rpm for 5 minutes. Finally the supernatant was collected and stored. The same method was used to produce more amounts of 2D nanomaterials of both $MoS_{2}$ and $WS_{2}$. The schematics of exfoliation and drop-casting can be found in supporting documents.

Before drop-casting it was useful to know the relative concentration of $MoS_{2}$ and $WS_{2}$. For this purpose, the absorbance of the two dispersions was measured for a few different wavelengths using a Perkin-Elmer Lambda 35 UV-vis-NIR Spectrometer. By adding some amounts of IPA into the denser dispersion the relative concentration of the two dispersions was equalized; this would make the gradual mixing task much easier since the goal was to create a vector of different combinations of these two materials by gradually changing the relative proportions, being 100\% $WS_{2}$ (for $f_{1}$), gradually adding $MoS_{2}$ and reducing $WS_{2}$ in steps of 10\% to create new combinations (for $f_{2},...,f_{10}$) and finally reaching to 100\% $MoS_{2}$ (for $f_{11}$). Altogether 11 of such combinations were made, stored in separate sealed containers and later were drop-casted using micro-liter onto surface of separate clean glass slides (\figref{gradelightT}(a)). The number of drops for each glass slide were kept the same to create almost the same thickness and area of drop-casted materials on glass. The IPA dried out in a few seconds. The slides with nanomaterials on them (called "filters") were annealed in nominal vacuum for 12 hours to stabilize them and eliminate any trace of IPA.

\textbf{Transmittance.} Since the glass itself was not part of the nanomaterial filters, by placing a clean glass slide as reference in reference beam position of UV-vis-NIR the effect of glass itself was removed and the outcome was transmittance of the 2D nanomaterials only (\figref{gradelightT}(b)). The transmittance spectrum of each filter was measured about 120 times over the 325--1100nm range with 1nm precision of UV-vis. \figref{gradelightT}(b) shows the transmittance spectrum of 11 filters where each curve is averaged from 120 measurements of the same filter on the scale of 0-1, with 1 being 100\% transparent. As it can be seen all filters have finite non-zero transmittance over a fairly large wavelength range.

\textbf{Atomic Force Microscopy (AFM), Scanning Electron Microscopy (SEM), and Ramman Spectrum}. In order to obtain the nanomaterial properties of the exfoliated TMDs, a single drop of $MoS_{2}$ and $WS_{2}$ was drop-casted on two separated silicon/silicon dioxide slides. The reason behind this was first, to do layer-thickness investigation via AFM since the glass slide does not posses as smooth surface as Silicon dioxide wafer does; second, Characteristic Ramman Peak of Silicon dioxide is a standard measure to study the nanomaterial properties. AFM and SEM investigations revealed that a typical nanoflake of $MoS_{2}$ was about 500nm long and 30nm thick. The SEM and AFM images with the corresponding line profiles of AFM'ed areas and Ramman spectrum of the exfoliated samples can be found in supporting documents section.

\textbf{Statistical Model; Generating Large Synthetic Training Samples for the Inference Model.} To calculate the individual filter likelihood $P(t_{k}\mid \lambda_{j})$, it was assumed that transmittance data $t_k$ of filter $k$ at wavelength $\lambda_{i}$ comes from a Gaussian normal distribution with the mean value of $\mu_{k}^{i}$ and standard deviation of $\sigma_{k}^{i}$, so 
$P(t_{k}\mid \lambda_{i})\sim \mathcal{N}(\mu_{k}^{i},\sigma_{k}^{i})$. This likelihood was used as a generative model to synthesize large amount of training samples, which were used in our training error reported in \figref{bayesianknn}(a) and \figref{entirebayesian}. For each wavelength using the mentioned likelihood, 100 vector of 11 elements (1 transmittance per each of 11 filters) were generated. The corresponding wavelength of each of these 100 synthesized examples per wavelength was estimated, and the 100 estimated wavelengths were averaged and used to find the average error. The same was performed for all wavelengths in the mentioned spectrum.
This is where the test samples were collected independent from the training data using UV-vis-NIR; in another word, the test samples were not part of training set seen by the code. For each test light, 100 test vectors of 11 elements (1 per filter) were sampled and collected. Before doing estimation, every 10 samples were averaged, which means only 10 vectors per test sample were obtained. The wavelength of these 10 (averaged) samples was estimated using Bayesian inference, and the average test errors are presented on the same plots as the training errors are. 

\textbf{k-Nearest Neighbor.} By averaging the 120 spectrum per filter, a single spectrum per filter was obtained which was a $775\times11$ matrix of transmittance values (11 filter and 775 spectrum elements between 325--1100nm). With this $T$ matrix at hand, the sum of squares of absolute errors between the test vector ($1\times11$ elements) and each row (wavelength) of $T$ was calculated. The wavelength with smallest square value was picked as the best estimation.

\section{Acknowledgment}
DH and SK acknowledges financial support from NSF ECCS 1351424, and a Northeastern University Provost's Tier 1 Interdisciplinary seed grant.

\section{Supporting Information Available}
\label{sec:supporting}
Supporting Information Available: 1. Fabrication of filters; 2. Atomic force microscopy (AFM) images and line-profile of the samples. 3. Scanning electron microscopy (SEM) images of the samples. Posterior probability estimation examples for test samples at some wavelengths. 4. Normalized root-mean-square error percentage (RMS\%) equation.
This material is available free of
charge via the Internet at http://pubs.acs.org.

\bibliographystyle{splncs}
\bibliography{paper}

\begin{thebibliography}{10}
\providecommand{\url}[1]{\texttt{#1}}
\providecommand{\urlprefix}{URL }
\providecommand{\doi}[1]{https://doi.org/#1}

\bibitem{altman1992introduction}
Altman, N.S.: An introduction to kernel and nearest-neighbor nonparametric
  regression. The American Statistician  \textbf{46}(3),  175--185 (1992)

\bibitem{an2013tunable}
An, X., Liu, F., Jung, Y.J., Kar, S.: Tunable graphene--silicon heterojunctions
  for ultrasensitive photodetection. Nano letters  \textbf{13}(3),  909--916
  (2013)

\bibitem{bassett2016maximum}
Bassett, R., Deride, J.: Maximum a posteriori estimators as a limit of bayes
  estimators. Mathematical Programming pp. 1--16 (2016)

\bibitem{berg2017layer}
Berg, M., Keyshar, K., Bilgin, I., Liu, F., Yamaguchi, H., Vajtai, R., Chan,
  C., Gupta, G., Kar, S., Ajayan, P., et~al.: Layer dependence of the
  electronic band alignment of few-layer mo s 2 on si o 2 measured using
  photoemission electron microscopy. Physical Review B  \textbf{95}(23),
  235406 (2017)

\bibitem{bernardo2001bayesian}
Bernardo, J.M., Smith, A.F.: Bayesian theory (2001)

\bibitem{boyer1987color}
Boyer, K.L., Kak, A.C.: Color-encoded structured light for rapid active
  ranging. IEEE Transactions on Pattern Analysis \& Machine Intelligence (1),
  14--28 (1987)

\bibitem{copeland2009color}
Copeland, K.G., Toerne, K.A.: Color infrared light sensor, camera, and method
  for capturing images (Jun~25 2009), uS Patent App. 11/960,302

\bibitem{cormen2001greedy}
Cormen, T.H., Leiserson, C.E., Rivest, R.L., Stein, C.: Greedy algorithms.
  Introduction to algorithms  \textbf{1} (2001)

\bibitem{de2018graphene}
De~Sanctis, A., Mehew, J., Craciun, M., Russo, S.: Graphene-based light
  sensing: fabrication, characterisation, physical properties and performance.
  Materials  \textbf{11}(9), ~1762 (2018)

\bibitem{decker1978optical}
Decker, D.L., Tolles, W.M.: Optical notch filter utilizing electric dipole
  resonance absorption (Jul~11 1978), uS Patent 4,099,854

\bibitem{goossen1994voltage}
Goossen, K.W.: Voltage-tunable photodetector (Jul~12 1994), uS Patent 5,329,136

\bibitem{goudjil1998photochromic}
Goudjil, K., Sandoval, R.: Photochromic ultraviolet light sensor and
  applications. Sensor Review  \textbf{18}(3),  176--177 (1998)

\bibitem{grigorenko2012graphene}
Grigorenko, A., Polini, M., Novoselov, K.: Graphene plasmonics. Nature
  photonics  \textbf{6}(11), ~749 (2012)

\bibitem{hu2018wavelength}
Hu, X., Liu, H., Wang, X., Zhang, X., Shan, Z., Zheng, W., Li, H., Wang, X.,
  Zhu, X., Jiang, Y., et~al.: Wavelength selective photodetectors integrated on
  a single composition-graded semiconductor nanowire. Advanced Optical
  Materials p. 1800293 (2018)

\bibitem{johnston2015optoelectronics}
Johnston, M.B.: Optoelectronics: colour-selective photodiodes. Nature Photonics
   \textbf{9}(10), ~634 (2015)

\bibitem{kim2014graphene}
Kim, J.T., Yu, Y.J., Choi, H., Choi, C.G.: Graphene-based plasmonic
  photodetector for photonic integrated circuits. Optics express
  \textbf{22}(1),  803--808 (2014)

\bibitem{kipping2011detection}
Kipping, D.M., Spiegel, D.S.: Detection of visible light from the darkest
  world. Monthly Notices of the Royal Astronomical Society: Letters
  \textbf{417}(1),  L88--L92 (2011)

\bibitem{knox1963theory}
Knox, R.S.: Theory of excitons. Solid State Phys.  \textbf{5} (1963)

\bibitem{konstantatos2006ultrasensitive}
Konstantatos, G., Howard, I., Fischer, A., Hoogland, S., Clifford, J., Klem,
  E., Levina, L., Sargent, E.H.: Ultrasensitive solution-cast quantum dot
  photodetectors. Nature  \textbf{442}(7099), ~180 (2006)

\bibitem{lai1994design}
Lai, K., Campbell, J.C.: Design of a tunable gaas/algaas multiple-quantum-well
  resonant-cavity photodetector. IEEE journal of quantum electronics
  \textbf{30}(1),  108--114 (1994)

\bibitem{lee1989bayesian}
Lee, P.M.: Bayesian statistics. Oxford University Press London: (1989)

\bibitem{lippke1981distribution}
Lippke, J.A., Gordon, L.K., Brash, D.E., Haseltine, W.A.: Distribution of uv
  light-induced damage in a defined sequence of human dna: detection of
  alkaline-sensitive lesions at pyrimidine nucleoside-cytidine sequences.
  Proceedings of the National Academy of Sciences  \textbf{78}(6),  3388--3392
  (1981)

\bibitem{mori2003ultraviolet}
Mori, K., Nishida, T., Hashimoto, H.: Ultraviolet light measuring chip and
  ultraviolet light sensor using the same (Apr~22 2003), uS Patent 6,551,493

\bibitem{nun2014supervised}
Nun, I., Pichara, K., Protopapas, P., Kim, D.W.: Supervised detection of
  anomalous light curves in massive astronomical catalogs. The Astrophysical
  Journal  \textbf{793}(1), ~23 (2014)

\bibitem{radisavljevic2011single}
Radisavljevic, B., Radenovic, A., Brivio, J., Giacometti, i.V., Kis, A.:
  Single-layer mos2 transistors. Nature nanotechnology  \textbf{6}(3), ~147
  (2011)

\bibitem{rieke2003detection}
Rieke, G.: Detection of Light: from the Ultraviolet to the Submillimeter.
  Cambridge University Press (2003)

\bibitem{sahami1996learning}
Sahami, M.: Learning limited dependence bayesian classifiers. In: KDD. vol.~96,
  pp. 335--338 (1996)

\bibitem{shenoi2005introduction}
Shenoi, B.A.: Introduction to digital signal processing and filter design. John
  Wiley \& Sons (2005)

\bibitem{splendiani2010emerging}
Splendiani, A., Sun, L., Zhang, Y., Li, T., Kim, J., Chim, C.Y., Galli, G.,
  Wang, F.: Emerging photoluminescence in monolayer mos2. Nano letters
  \textbf{10}(4),  1271--1275 (2010)

\bibitem{tang2008nanometre}
Tang, L., Kocabas, S.E., Latif, S., Okyay, A.K., Ly-Gagnon, D.S., Saraswat,
  K.C., Miller, D.A.: Nanometre-scale germanium photodetector enhanced by a
  near-infrared dipole antenna. Nature Photonics  \textbf{2}(4), ~226 (2008)

\bibitem{wang2014bionic}
Wang, X., Xu, L., Sun, H., Xin, J., Zheng, N.: Bionic vision inspired on-road
  obstacle detection and tracking using radar and visual information. In:
  Intelligent Transportation Systems (ITSC), 2014 IEEE 17th International
  Conference on. pp. 39--44. IEEE (2014)

\bibitem{wilke2010stimulation}
Wilke, R.G., Moghaddam, G.K., Dokos, S., Suaning, G., Lovell, N.H.: Stimulation
  of the retinal network in bionic vision devices: From multi-electrode arrays
  to pixelated vision. In: International Conference on Neural Information
  Processing. pp. 140--147. Springer (2010)

\bibitem{wilson1969transition}
Wilson, J.A., Yoffe, A.: The transition metal dichalcogenides discussion and
  interpretation of the observed optical, electrical and structural properties.
  Advances in Physics  \textbf{18}(73),  193--335 (1969)

\bibitem{wood1985wavelength}
Wood, T., Burrus, C., Gnauck, A., Wiesenfeld, J., Miller, D., Chemla, D.,
  Damen, T.: Wavelength-selective voltage-tunable photodetector made from
  multiple quantum wells. Applied Physics Letters  \textbf{47}(3),  190--192
  (1985)

\bibitem{xia2009ultrafast}
Xia, F., Mueller, T., Lin, Y.m., Valdes-Garcia, A., Avouris, P.: Ultrafast
  graphene photodetector. Nature nanotechnology  \textbf{4}(12), ~839 (2009)

\end{thebibliography}
\end{document}